\documentclass[aps,pre,twocolumn,superscriptaddress,longbibliography]{revtex4-2}

\usepackage{amsmath,amssymb,amsfonts}
\usepackage{graphicx}
\usepackage{bm}
\usepackage{hyperref}
\usepackage{xcolor}
\usepackage{mathtools}
\usepackage{subcaption}

\renewcommand{\vr}{\mathbf{r}}
\newcommand{\vt}{\hat{\mathbf{t}}}
\newcommand{\kB}{k_{\mathrm{B}}}
\newcommand{\Mh}{M_{\mathrm{h}}}
\newcommand{\kb}{\kappa_{\mathrm{b}}}
\newcommand{\ks}{\kappa_{\mathrm{s}}}
\newcommand{\Fact}{F_{\mathrm{act}}}
\newcommand{\Pe}{\mathrm{Pe}}
\newcommand{\MSD}{\langle \delta^2(\tau) \rangle}
\newcommand{\Ct}{C_t(\tau)}
\newcommand{\Cs}{C_s(s)}
\newcommand{\Nf}{N_{\mathrm{f}}}
\newcommand{\td}{\tau_{\mathrm{d}}}
\newcommand{\ti}{\tau_{\mathrm{i}}}
\newcommand{\ta}{\tau_{\mathrm{a}}}

\begin{document}

\title{Multi-filament coordination rescues active transport from inertia-induced spinning arrest}

\author{Anuradha Rajput}
\affiliation{School of Computational and Integrative Sciences, Jawaharlal Nehru University, New Delhi 110067, India}

\author{Arnab Bhattacharjee}
\email{arnab@jnu.ac.in}
\affiliation{School of Computational and Integrative Sciences, Jawaharlal Nehru University, New Delhi 110067, India}

\author{Annwesha Dutta}
\email{annwesha@jnu.ac.in}
\affiliation{School of Computational and Integrative Sciences, Jawaharlal Nehru University, New Delhi 110067, India}

\date{\today}

\begin{abstract}
Active filaments driven by tangential forces can become trapped in a spinning state when attached to a heavy head, where activity and inertia drive persistent rotation rather than directed transport.  Using three-dimensional Langevin dynamics of tangentially driven bead-spring chains anchored to a common heavy head, we demonstrate that increasing the filament number $\Nf$ systematically \emph{rescues} directed transport by sterically preventing the coiled conformations that underlie spinning.  The rescue is established through three independent diagnostics: (i)~the mean-square displacement recovers monotonic growth (transport rescue), (ii)~the spatial tangent autocorrelation loses its negative dip signaling helical coiling (conformational rescue), and (iii)~the tangent time autocorrelation ceases crossing zero (orientational rescue).  At high bending stiffness ($\kb = 1000$), coiling is fully eliminated at a critical filament number $\Nf^* \approx 3$.  At moderate stiffness ($\kb = 100$), residual coiling persists ($\min C_s \approx -0.13$) yet transport is still rescued---demonstrating that the destruction of spinning \emph{coherence}, not coiling elimination, is the essential mechanism.  The multi-filament architecture achieves up to five orders of magnitude transport enhancement.  Two physically distinct rescue pathways emerge: at high stiffness, steric constraints force filaments into a coordinated bundle sustaining directed propulsion; at low stiffness, steric interactions destroy orientational coherence, producing enhanced active diffusion.  These results demonstrate a purely mechanical, density-independent route to overcome inertia-induced motility arrest, with implications for synthetic microswimmer design, motor-driven filament assays, and multi-filament organization in biological systems.
\end{abstract}


\maketitle

\section{Introduction}
\label{sec:intro}

Active filaments---polymers driven out of equilibrium by internal or external forces---are ubiquitous across biological scales.  Motor-driven cytoskeletal filaments form the basis of intracellular transport and cell motility~\cite{Winkler2020,Eisenstecken2016poly,Ghosh2014,Brangwynne2008}, bacterial flagella propel cells through viscous environments~\cite{Patteson2015,Eisenstecken2016,Kumar2009,Turner2000,Berg2003}, and cilia generate fluid flows through synchronized beating~\cite{Elgeti2013,Gilpin2020}.  These systems operate far from equilibrium, and their motility emerges from a complex interplay of activity, elasticity, dissipation, and inertia~\cite{Bar2020,Vliegenthart2020,Yadav2024,Zhu2024,IseleHolder2015,Ramaswamy2010,Marchetti2013}.  In the overdamped regime, tangentially self-propelled semiflexible polymers exhibit states ranging from open beating to compact spirals~\cite{IseleHolder2015,Winkler2020,Bianco2018}, with activity producing enhanced diffusion~\cite{Ghosh2014,Chelakkot2014}, spiral formation~\cite{IseleHolder2015,Jiang2014}, activity-dependent persistence length~\cite{Eisenstecken2016poly,Anand2019}, and conformational collapse~\cite{Sarkar2017}.

A key development has been the recognition that \emph{inertia} qualitatively alters active filament dynamics.  While classical active matter theory operates in the overdamped regime~\cite{Ramaswamy2010,Marchetti2013}, underdamped dynamics introduce new phenomena: L\"owen~\cite{Lowen2020} showed that finite mass modifies the ballistic-to-diffusive crossover, Scholz \textit{et al.}~\cite{Scholz2018} demonstrated inertial delay effects, and Caprini \textit{et al.}~\cite{Caprini2021} showed that inertia modifies active phase behavior.  For active filaments specifically, Isele-Holder \textit{et al.}~\cite{IseleHolder2015} identified spontaneous spiral formation, and Karan \textit{et al.}~\cite{Karan2024} mapped a re-entrant phase diagram featuring open, beating, and spinning-spiral states as a function of inertia and activity.  Khalilian \textit{et al.}~\cite{Khalilian2024} analyzed the spiral onset as a subcritical pitchfork bifurcation, Fazelzadeh \textit{et al.}~\cite{Fazelzadeh2023} showed that inertia enhances effective persistence length, and Tejedor \textit{et al.}~\cite{Tejedor2024} demonstrated inertia-strengthened head--tail asymmetry.  Further studies have characterized inertial effects on polymer relaxation~\cite{vanSteijn2024}, active nematics~\cite{Mokhtari2019}, crowded transport~\cite{Liu2020}, and dynamical scaling~\cite{MartinRoca2024}.  Complementary work on asymmetric activity distributions~\cite{Li2023,Hu2022,Wu2022,Wang2022} has shown that concentrating forces---whether active or inertial---at specific locations along the backbone profoundly affects the motility mode.

Taken together, these findings establish that spinning arrest---persistent rotation rather than translation---is a \emph{generic} feature of inertial active filament dynamics.  The central question motivating this work is: \emph{can this motility arrest be overcome, and if so, by what mechanism?}

In biology and synthetic systems, active filaments are often organized into multi-filament architectures: flagellar bundles on bacterial cell bodies~\cite{Namba1989,Turner2000,Berg2003,Macnab1977}, actin stress fibers at focal adhesions~\cite{Tojkander2012}, microtubule asters at centrosomes~\cite{Brinkley1985}, and interacting filaments in motility assays~\cite{Schaller2010,Sumino2012}.  Whether such organization provides a functional advantage \emph{specifically by suppressing spinning arrest} has not been investigated.

Two prior studies are particularly relevant.  Janzen and Matoz-Fernandez~\cite{Janzen2024} showed that in dense two-dimensional suspensions, inter-filament collisions destabilize spirals through a density-dependent reentrant transition---demonstrating that steric interactions \emph{can} disrupt spinning, but requiring high density.  Buglakov \textit{et al.}~\cite{Buglakov2025} studied active star polymers with correlated activity, finding metastable coordinated states, but did not address spinning suppression or inertia.  Related work has examined passive star polymers~\cite{Grest1987,Vlassopoulos2014}, clamped active filaments~\cite{Chelakkot2021}, active ring polymers~\cite{Vatin2024}, and confined active chains~\cite{Anand2019}.

In this work, we demonstrate that anchoring multiple active filaments to a common heavy head provides a \emph{purely architectural} route to overcome spinning arrest through steric frustration of the coiled conformation, independent of suspension density.  We characterize the rescue through three complementary diagnostics---mean-square displacement, spatial tangent autocorrelation, and tangent time autocorrelation---that independently identify the critical filament number and reveal two qualitatively distinct rescue mechanisms at low versus high bending stiffness.

The remainder of this paper is organized as follows.  Section~\ref{sec:model} describes the model and simulation methodology.  Section~\ref{sec:observables} defines the observables.  Section~\ref{sec:spinning} establishes the spinning arrest in single filaments.  Section~\ref{sec:rescue} presents the multi-filament rescue results.  We discuss implications and conclude in Sec.~\ref{sec:discussion}.

\section{Model and simulation}
\label{sec:model}

\subsection{Active filament as a bead-spring chain}

\begin{figure}[t]
     \centering
     \begin{subfigure}[b]{0.4\textwidth}
         \centering
         \includegraphics[width=\columnwidth]{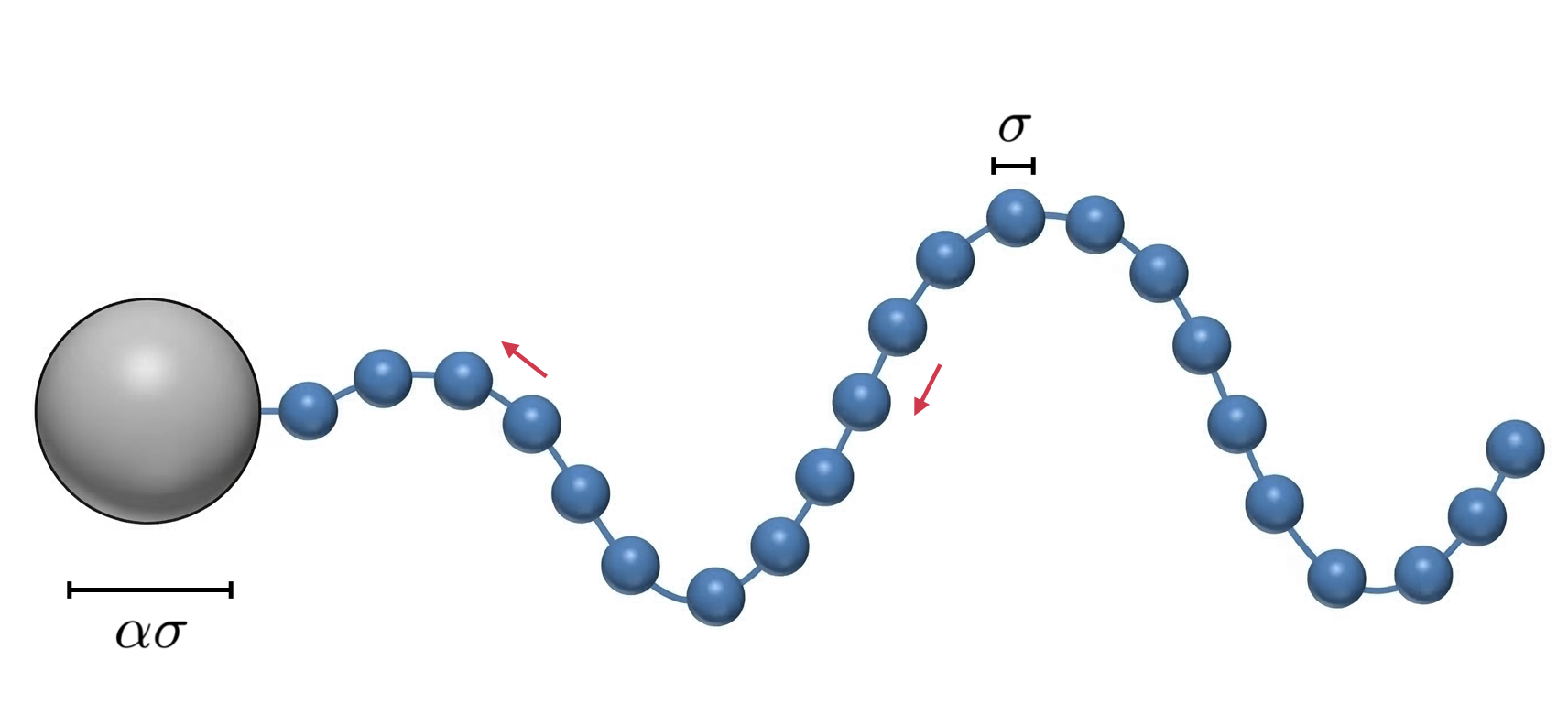}
         \caption{Single chain}
         \label{fig:single_chain}
     \end{subfigure}
     \hfill
     \begin{subfigure}[b]{0.3\textwidth}
         \centering
         \includegraphics[width=\columnwidth]{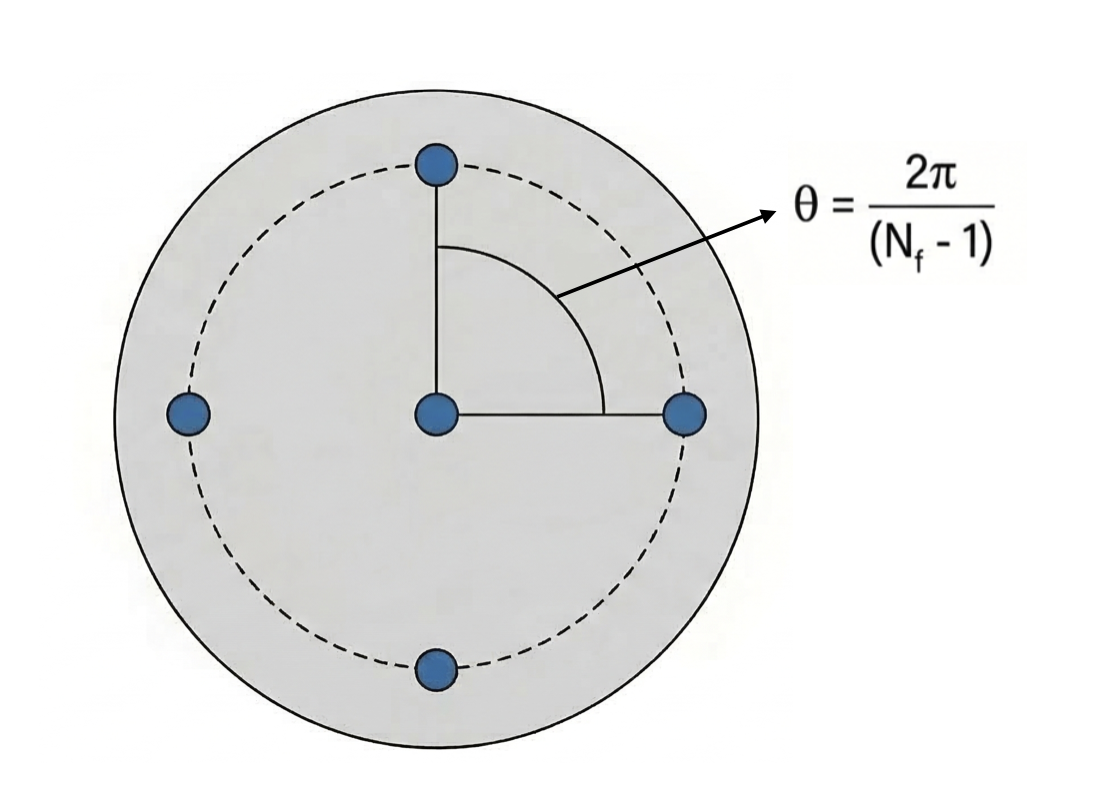}
         \caption{Multi-filament arrangement}
         \label{fig:flagella}
     \end{subfigure}
     \caption{Schematic of the active filament model. (a)~Single filament ($\Nf{=}1$): head bead (diameter $\alpha\sigma$) followed by $N$ body beads with tangential active forces directed toward the head. (b)~Multi-filament arrangement ($\Nf > 1$): $\Nf$ chains anchored to a common head.}
     \label{fig:schematic_main}
\end{figure}

We model an active filament as a bead-spring chain of $N{+}1$ monomers in three dimensions (Fig.~\ref{fig:schematic_main}).  A head bead ($i{=}1$) of diameter $\alpha\sigma$ ($\alpha > 1$) is followed by $N$ body beads of diameter $\sigma$.  The head bead serves as a proxy for a cargo, cell body, or other heavy structure to which the filament is anchored, and its larger mass and size relative to the body beads introduces the inertial asymmetry that drives the spinning transition.

The underdamped Langevin equation governing the motion of the $i$th bead is
\begin{equation}
m_i \, \ddot{\vr}_i = -\gamma_i \, \dot{\vr}_i - \nabla_i U + f_a^{(i)} \, \vt_i + \sqrt{2\gamma_i \kB T}\; \bm{\eta}_i(t),
\label{eq:langevin}
\end{equation}
where $m_i$ is the mass, $\gamma_i$ is the friction coefficient, and $f_a^{(i)} = f_a$ for body beads ($i = 2, \ldots, N{+}1$) and $f_a^{(1)} = 0$ for the head bead.  The total conservative potential $U = U_{\mathrm{bond}} + U_{\mathrm{bend}} + U_{\mathrm{WCA}}$ comprises three contributions.

The \emph{bond potential} enforces chain connectivity via a harmonic spring:
\begin{equation}
U_{\mathrm{bond}} = \frac{1}{2}\ks \sum_{i} \left(r_{i,i+1} - r_{0,i}\right)^{2},
\label{eq:bond}
\end{equation}
where $\ks$ is the spring constant, $r_{i,i+1} = |\vr_{i+1} - \vr_i|$ is the distance between consecutive beads, and $r_{0,i} = \mathrm{rad}_i + \mathrm{rad}_{i+1}$ is the equilibrium bond length (sum of bead radii).  We use $\ks = 1000\,\kB T/\sigma^2$ to enforce near-rigid bonds.  The equilibrium bond length $r_{0,i}$ equals the sum of the radii of the two bonded beads: $r_0 = \sigma$ for body--body bonds and $r_0 = (1+\alpha)\sigma/2$ for head--body bonds.

The \emph{bending potential} imposes semiflexibility through a discrete bond-difference form:
\begin{equation}
U_{\mathrm{bend}} = \frac{\kb}{4} \sum_{i} \left|\vt_i - \vt_{i+1}\right|^{2} = \frac{\kb}{2} \sum_{i} \left(1 - \cos\theta_i\right),
\label{eq:bend}
\end{equation}
where $\vt_i$ is the unit bond vector between beads $i$ and $i{+}1$, $\theta_i$ is the angle between consecutive bond vectors, and $\kb$ is the bending modulus.  The persistence length in the passive limit is $\ell_p = (\kb/2)\sigma/\kB T$, reflecting the factor of $1/2$ relative to the standard Kratky--Porod form~\cite{Doi1986}.  We explore bending moduli ranging from $\kb = 0$ (fully flexible) to $\kb = 1000\,\kB T$.

The \emph{excluded-volume interaction} is modeled by the Weeks--Chandler--Andersen (WCA) potential~\cite{Weeks1971,Kremer1990}:
\begin{equation}
U_{\mathrm{WCA}}(r_{ij}) = 
\begin{cases}
4\epsilon\!\left[\left(\frac{d_{ij}}{r_{ij}}\right)^{\!12} - \left(\frac{d_{ij}}{r_{ij}}\right)^{\!6}\right] + \epsilon, & r_{ij} < 2^{1/6}d_{ij}, \\[4pt]
0, & r_{ij} \geq 2^{1/6}d_{ij},
\end{cases}
\label{eq:wca}
\end{equation}
where $r_{ij} = |\vr_i - \vr_j|$, $d_{ij} = (\sigma_i + \sigma_j)/2$ is the interaction range set by the bead diameters, and $\epsilon = \kB T$ sets the energy scale.  The WCA potential acts between all bead pairs separated by more than two bonds along the chain (i.e., bonded and 1--3 pairs are excluded), including beads on different filaments in the multi-filament model.  This purely repulsive interaction is the sole source of inter-filament coupling and provides the steric frustration mechanism central to our results.

The active force on each body bead is $f_a \vt_i$, where $f_a$ is the constant active force magnitude and $\vt_i = (\vr_{i-1} - \vr_i)/|\vr_{i-1} - \vr_i|$ is the unit tangent vector directed from bead $i$ toward bead $i{-}1$ (i.e., toward the head)~\cite{IseleHolder2015,Kolomeisky2007}.  This tangential driving mimics the force generation by molecular motors walking along cytoskeletal filaments.  The head bead ($i{=}1$) does not carry an active force; it is propelled passively by the cumulative thrust of the body beads.

The stochastic term $\bm{\eta}_i(t)$ is a three-dimensional Gaussian white noise satisfying $\langle \eta_i^\mu(t) \rangle = 0$ and $\langle \eta_i^\mu(t)\,\eta_j^\nu(t') \rangle = \delta_{ij}\delta_{\mu\nu}\delta(t - t')$, where $\mu, \nu \in \{x, y, z\}$.  The noise amplitude $\sqrt{2\gamma_i \kB T}$ ensures that the fluctuation-dissipation relation is satisfied in the absence of activity ($f_a = 0$), recovering the correct equilibrium Boltzmann distribution.

\subsection{Non-dimensionalization}
\label{sec:nondim}

The system dynamics are governed by three fundamental timescales: the inertial relaxation time $\ti = m/\gamma$, the Brownian diffusion time $\td = \gamma\sigma^2/\kB T$, and the active time $\ta = \sigma/v_0$ (where $v_0 = f_a/\gamma$ is the free-swimming speed of a body bead).  We non-dimensionalize Eq.~\eqref{eq:langevin} using $\sigma$ for length, $\td$ for time, and $\kB T$ for energy.  Defining $\tilde{\vr}_i = \vr_i/\sigma$ and $\tilde{t} = t/\td$, and dividing through by $\kB T/\sigma$, the dimensionless equation for body beads ($m_i = m$, $\gamma_i = \gamma$) becomes
\begin{equation}
\mathcal{M}\,\ddot{\tilde{\vr}}_i
= -\dot{\tilde{\vr}}_i
  - \tilde{\nabla}_i \tilde{U}
  + \Pe\,\vt_i
  + \sqrt{2}\;\tilde{\bm{\eta}}_i,
\label{eq:nondim}
\end{equation}
where the two key dimensionless parameters are
\begin{equation}
\mathcal{M} = \frac{m\kB T}{\gamma^2\sigma^2} = \frac{\ti}{\td}, \qquad \Pe = \frac{\sigma f_a}{\kB T}.
\end{equation}
The dimensionless mass $\mathcal{M}$ measures the ratio of inertial to diffusive timescales: $\mathcal{M} \ll 1$ corresponds to the overdamped regime, while $\mathcal{M} \gtrsim 1$ signals significant inertial effects.  The P\'eclet number $\Pe$ measures the ratio of active to thermal forces.

For the head bead, with mass $m_\mathrm{h} = \alpha^3 m$ (volume scaling at constant density) and friction $\gamma_\mathrm{h} = \alpha\gamma$ (Stokes scaling with diameter), the dimensionless head mass is defined analogously as
\begin{equation}
\Mh = \frac{m_\mathrm{h}\kB T}{\gamma_\mathrm{h}^2\sigma^2} = \frac{\alpha^3 m \kB T}{\alpha^2 \gamma^2\sigma^2} = \alpha\mathcal{M}.
\label{eq:Mh}
\end{equation}
The linear scaling $\Mh = \alpha\mathcal{M}$ arises because the Stokes friction ($\gamma_\mathrm{h} \propto \alpha$) partially compensates the volumetric mass ($m_\mathrm{h} \propto \alpha^3$).  By contrast, non-dimensionalizing with the active time $\ta$ gives a head mass $\alpha^2\mathcal{M}'$ that couples geometry and inertia more strongly.  Our choice of $\td$ as the reference cleanly separates the inertial ($\mathcal{M}$), active ($\Pe$), and geometric ($\alpha$) control parameters (see Appendix~\ref{app:nondim} for the full derivation and comparison).

\subsection{Multi-filament model}
\label{sec:multi}

The multi-filament system consists of $\Nf$ active filaments anchored to a common head bead (Fig.~\ref{fig:schematic_main}).  Each filament is an independent chain of $N$ body beads obeying Eq.~\eqref{eq:nondim}.  The first body bead of each filament is bonded to the shared head bead, with its center placed at contact distance $r_0 = (\alpha+1)\sigma/2$ from the head center.

For $\Nf = 1$ the single filament is attached at the north pole of the head bead and initially extends along the $+z$ axis.  For $\Nf > 1$ the first filament retains this axial placement, while the remaining $\Nf - 1$ filaments are distributed on an equatorial ring around the head with equal angular spacing $\theta = 2\pi/(\Nf - 1)$.  The anchor point of the $k$th equatorial filament ($k = 1, \ldots, \Nf{-}1$) lies in the $xy$-plane at
\begin{equation}
\vr_{\mathrm{anchor},k} = \bigl(r_0\cos[(k{-}1)\theta],\; r_0\sin[(k{-}1)\theta],\; 0\bigr).
\end{equation}
All filaments are initialized as straight rods extending along the $+z$ direction from their respective anchor points.  The passive chain ($f_a = 0$) is then equilibrated before switching on the active force, allowing the initial geometry to relax before data collection begins.

The head bead experiences the sum of all forces transmitted from all $\Nf$ filaments, as well as its own friction and thermal noise.  Using $m_\mathrm{h} = \alpha^3 m$ and $\gamma_\mathrm{h} = \alpha\gamma$, the dimensionless head equation is
\begin{equation}
\alpha\mathcal{M}\,\ddot{\tilde{\vr}}_\mathrm{h} = -\alpha\,\dot{\tilde{\vr}}_\mathrm{h} - \sum_{k=1}^{\Nf}\tilde{\nabla}_\mathrm{h} \tilde{U}_k + \sqrt{2\alpha}\;\tilde{\bm{\eta}}_\mathrm{h},
\label{eq:head}
\end{equation}
where the inertial prefactor $\alpha\mathcal{M} = \Mh$ [Eq.~\eqref{eq:Mh}], the friction prefactor is $\gamma_\mathrm{h}/\gamma = \alpha$, the noise amplitude is $\sqrt{2\gamma_\mathrm{h}/\gamma} = \sqrt{2\alpha}$, and the sum runs over all filaments $k$.  The head does not carry an active force.

Inter-filament interactions are \emph{purely steric}: beads on different filaments interact only through the WCA excluded-volume potential, Eq.~\eqref{eq:wca}.  No hydrodynamic coupling is included.  This choice is deliberate: by excluding fluid-mediated interactions, we isolate the role of steric and elastic coordination mechanisms in the rescue~\cite{Reigh2012,Kim2003}.  Hydrodynamic interactions would add an additional coupling channel that could enhance---but not diminish---the rescue effect; our results thus establish a lower bound on the multi-filament rescue.

\subsection{Simulation details}
\label{sec:simdetails}

We integrate the equations of motion using the Gr\o{}nbech-Jensen--Farago (GJF) Langevin integrator~\cite{GJF2013,Allen2017} with a dimensionless timestep $\Delta\tilde{t} = 10^{-3}$.  The GJF scheme preserves the correct configurational temperature and diffusion coefficient for any timestep at which the trajectory remains stable, making it well suited for underdamped Langevin dynamics across a wide range of friction coefficients~\cite{GJF2013}.  Each filament comprises $N = 100$ body beads, giving a total contour length of $L = 100\sigma$.  Simulations are run for $5 \times 10^6$ timesteps (corresponding to a total dimensionless time of $\tilde{t}_{\max} = 5000$), and each parameter set is averaged over $50$ independent runs with different random initial configurations and noise realizations.

We systematically explore the following parameter ranges: dimensionless head mass $\Mh = 1$--$40$; bending stiffness $\kb = 0$, $10$, $100$, $1000$; active force magnitude $\Fact = 4$, $10$, $20$, $40$; and number of filaments $\Nf = 1$--$7$.  The head-to-body size ratio is $\alpha = 3$ throughout, giving $\Mh = 3\mathcal{M}$.  Initial configurations are generated by placing beads along a straight line with small random perturbations and equilibrating the passive chain ($f_a = 0$) before switching on the active force.

\section{Observables}
\label{sec:observables}

We employ four complementary observables to characterize the spinning arrest and its rescue.  The first (MSD) probes transport directly; the second ($\Phi$) provides a binary classification of the dynamical state; the third (SACF) reveals the spatial conformational structure; and the fourth (TTAF) captures the temporal orientational dynamics.  Their quantitative agreement on the critical filament number $\Nf^*$ provides a stringent self-consistency test.

\subsection{Mean-square displacement and transport measure}
\label{sec:msd_obs}

The mean-square displacement of the head bead is computed as
\begin{equation}
\MSD = \left\langle |\vr_\mathrm{h}(t_0 + \tau) - \vr_\mathrm{h}(t_0)|^2 \right\rangle,
\end{equation}
where the average $\langle \cdot \rangle$ is taken over time origins $t_0$ within each trajectory (standard time-averaging) and then over all $50$ independent trajectories.  We track the head bead displacement because the head represents the biologically relevant cargo-carrying element.

To quantify transport efficiency with a single number that avoids the ambiguities of fitting power-law exponents over an arbitrary time window, we define the reference MSD:
\begin{equation}
\mathrm{MSD}_{\mathrm{ref}} = \MSD\big|_{\tau=\tau_\mathrm{ref}},
\end{equation}
where $\tau_\mathrm{ref}$ is chosen to be well beyond both the spinning period ($T_\mathrm{spin} \sim 60$--$90$) and the single-filament spinning coherence time ($\tau_\mathrm{spin} \sim 5000$--$15{,}000$), so that $\mathrm{MSD}_{\mathrm{ref}}$ captures the long-time transport behavior.

\subsection{Spinning order parameter $\Phi$}
\label{sec:phi_obs}

The spinning state produces a distinctive non-monotonic MSD: after an initial ballistic increase, the MSD \emph{decreases} at intermediate times as the filament completes half a rotation and the head returns toward its starting position, before increasing again as the slow radial drift accumulates over many cycles.  To detect this signature quantitatively, we compute the local log-log slope of the MSD:
\begin{equation}
\beta(\tau) = \frac{d\ln\MSD}{d\ln\tau},
\end{equation}
evaluated via centered finite differences on the logarithmic MSD with Savitzky--Golay smoothing (window~31, polynomial order~2) to suppress statistical noise.  Normal transport ($\MSD \sim \tau^\nu$) gives $\beta = \nu > 0$; the spinning state uniquely produces intervals where $\beta < 0$ (MSD decreasing).

The spinning order parameter is then defined as
\begin{equation}
\Phi = \frac{1}{\tau_2{-}\tau_1}\int_{\tau_1}^{\tau_2} \Theta\!\bigl(-\beta(\tau)\bigr)\,d\tau,
\label{eq:phi}
\end{equation}
where $\Theta$ is the Heaviside step function.  The integration window $[\tau_1, \tau_2] = [10, 10^3]$ is chosen to bracket the intermediate-time regime: $\tau_1 = 10$ excludes the short-time ballistic regime (where $\beta \approx 2$ regardless of spinning), and $\tau_2 = 10^3$ captures the MSD dips produced by coherent rotation while remaining below the coherence time where thermal drift dominates.  Physically, $\Phi$ measures the fraction of this window where the MSD is decreasing---a behavior unique to coherent periodic rotation.  We classify states with $\Phi > 0.1$ as spinning (the threshold excludes minor statistical fluctuations around $\beta = 0$) and $\Phi = 0$ as beating or propulsive.

\subsection{Spatial tangent autocorrelation (SACF)}
\label{sec:sacf_obs}

The bond tangent vector is defined consistently with the active force direction as $\vt_i = (\vr_{i-1} - \vr_i)/|\vr_{i-1} - \vr_i|$, pointing from bead $i$ toward the head.  The spatial tangent autocorrelation function
\begin{equation}
\Cs = \langle \vt_i(t) \cdot \vt_{i+s}(t) \rangle
\end{equation}
measures orientational correlations along the chain contour at separation $s$ (in bond units).  The average is taken over all body-bead indices $i$ (excluding the head), over time, and over all independent trajectories.  In the multi-filament case, $\Cs$ is computed separately for each filament and then averaged over filaments.

For a passive worm-like chain in equilibrium, $\Cs$ decays exponentially as $\exp(-s\sigma/\ell_p)$, remaining positive for all $s$.  The appearance of a \emph{negative} region in $\Cs$ at finite $s$ is the structural hallmark of helical coiling: it indicates that bonds separated by $s$ along the backbone are oriented in approximately opposite directions, meaning the chain has doubled back on itself.  The depth of the negative minimum (denoted $\min \Cs$) quantifies the degree of coiling, while its position along $s$ reflects the characteristic coiling wavelength (approximately half the coil circumference).

\subsection{Tangent time autocorrelation (TTAF)}
\label{sec:ttaf_obs}

The tangent time autocorrelation function
\begin{equation}
\Ct = \langle \vt_i(t_0) \cdot \vt_i(t_0 + \tau) \rangle
\end{equation}
measures how a given bond's orientation decorrelates over lag time $\tau$.  The average is taken over all body beads $i$, over time origins $t_0$, and over trajectories.  In the multi-filament case, the TTAF is computed separately for each filament to reveal inter-filament differences in orientational dynamics (see Sec.~\ref{sec:ttaf_results}).

For a passive polymer, $\Ct$ decays monotonically from unity to zero on the longest relaxation timescale.  For a beating (non-spinning) active filament, $\Ct$ also decays monotonically but with an activity-modified rate.  For spinning filaments, the TTAF exhibits qualitatively distinct behavior: it oscillates through zero with a well-defined period $T_{\mathrm{spin}}$ equal to the spinning rotation period, reflecting the periodic reversal of bond orientations as the coiled chain rotates.  The decay envelope of these oscillations defines the spinning coherence time $\tau_{\mathrm{spin}}$---the duration over which spinning maintains phase coherence before thermal fluctuations randomize it.  The TTAF thus provides both the spinning frequency (from $T_{\mathrm{spin}}$) and the spinning coherence time (from the envelope decay rate).

\section{Spinning arrest in single filaments}
\label{sec:spinning}

\begin{figure}[t]
    \centering
    \begin{subfigure}[b]{0.45\columnwidth}
        \centering
        \includegraphics[width=0.9\columnwidth]{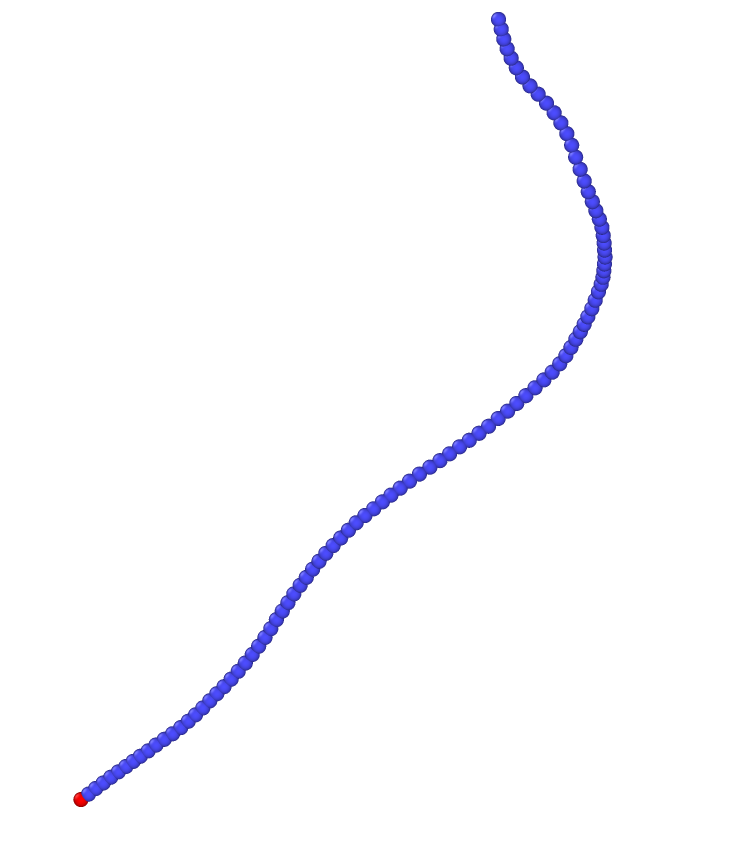}
        \caption{Beating conformation ($\Mh{=}10$)}
        \label{fig:beating}
    \end{subfigure}
    \hfill
    \begin{subfigure}[b]{0.45\columnwidth}
        \centering
        \includegraphics[width=0.9\columnwidth]{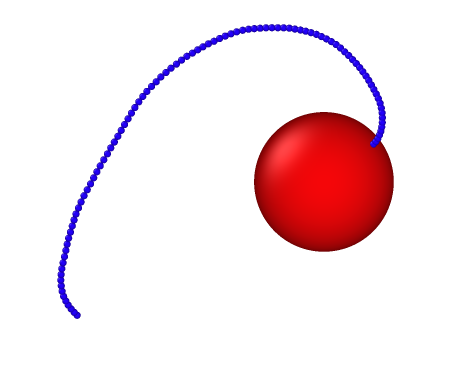}
        \caption{Spinning conformation ($\Mh{=}30$)}
        \label{fig:spinning_conf_b}
    \end{subfigure}
    
    \caption{Spinning arrest in a single filament ($\Nf{=}1$). (a)~Beating conformation ($\Mh{=}10$): extended, undulating shape producing directed transport. (b)~Spinning conformation ($\Mh{=}30$): filament coils tightly around the heavy head and rotates in place. Both at $\kb{=}1000$, $\Fact{=}40$.}
    \label{fig:spinning_conf}
\end{figure}

Consistent with previous studies~\cite{Karan2024,IseleHolder2015,Khalilian2024,Fazelzadeh2023}, we observe that heavy heads ($\Mh \gtrsim 10$--$20$ depending on $\kb$ and $\Fact$) induce a spinning state [Fig.~\ref{fig:spinning_conf}(b)] where the filament coils around the head and executes persistent rotation with strongly suppressed center-of-mass transport.  The mechanism is intuitive: the massive head resists the rapidly changing thrust direction, so the body beads sweep around it, forming a rotating helical coil.  Light heads translate smoothly, producing flagellar-like beating with efficient propulsion [Fig.~\ref{fig:spinning_conf}(a)].

\begin{figure}[t]
  \centering
  \includegraphics[width=0.9\columnwidth]{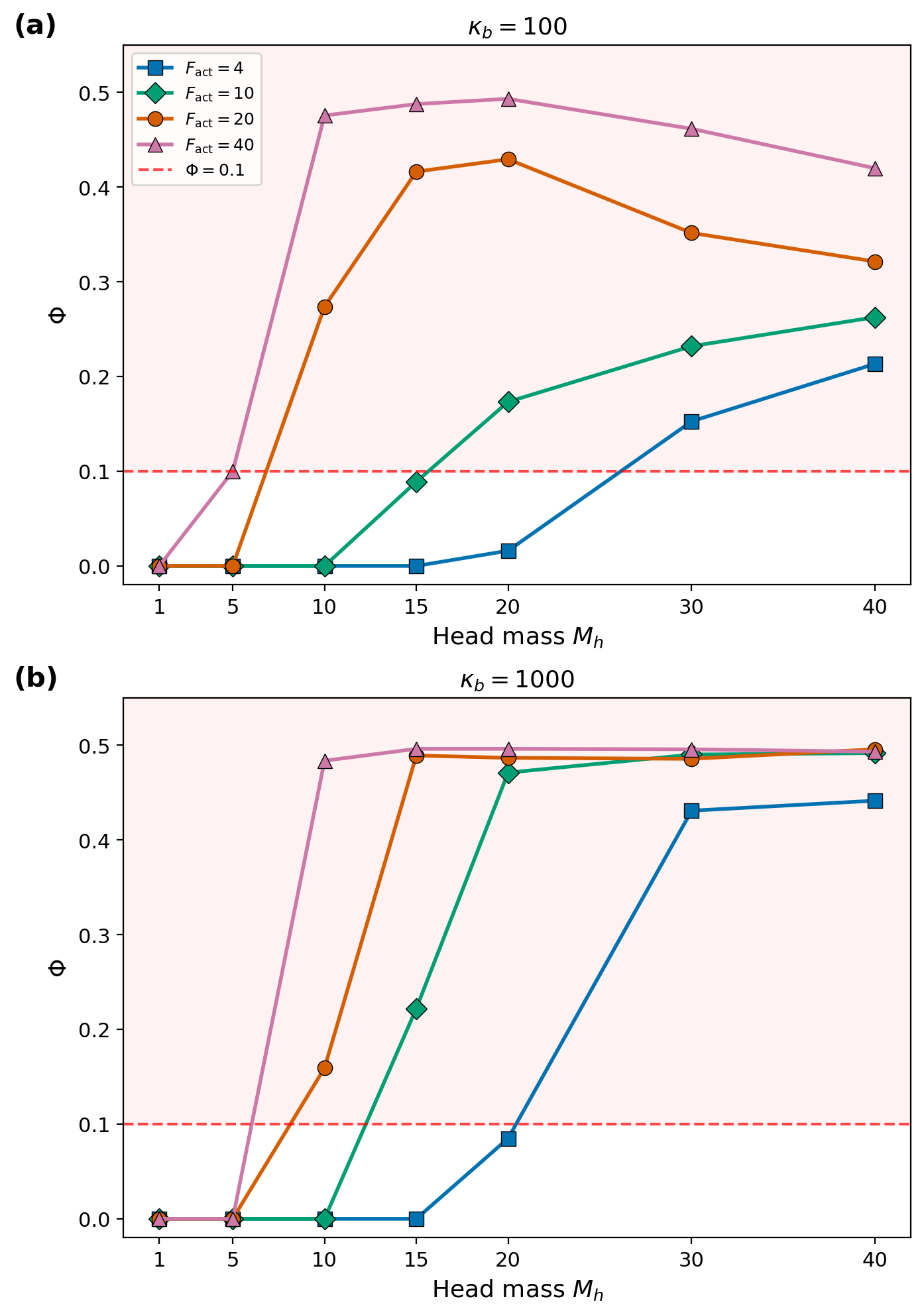}
  \caption{Phase diagram: $\Phi$ vs.\ $\Mh$ at $\kb = 100$ (blue) and $1000$ (orange) for $\Fact = 4,\,10,\,20,\,40$.  Dashed line: $\Phi = 0.1$ threshold.  Increasing activity shifts the spinning boundary to lower $\Mh$.}
  \label{fig:phase_diagram}
\end{figure}

The order parameter $\Phi$ maps the phase boundary quantitatively (Fig.~\ref{fig:phase_diagram}).  Two features are notable.  First, increasing activity \emph{promotes} spinning: the critical head mass decreases with $\Fact$ because stronger active forces generate larger torques around the heavy head (at $\Fact = 4$, spinning requires $\Mh \geq 20$; at $\Fact = 40$, even $\Mh = 5$ spins).  Second, bending stiffness amplifies spinning by suppressing the traveling-wave undulations needed for propulsion.

\begin{figure}[t]
  \centering
  \includegraphics[width=0.95\columnwidth]{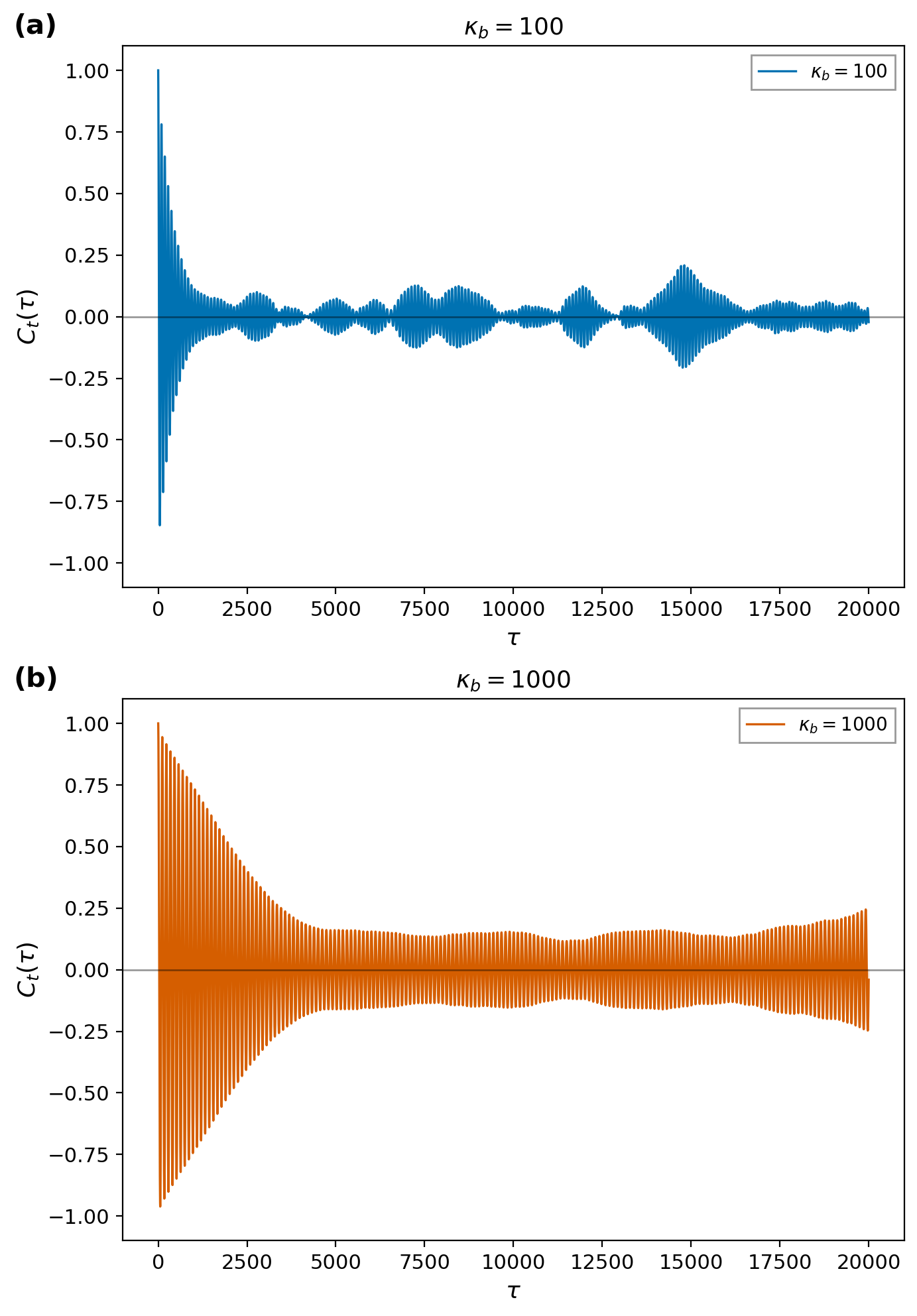} 
  \caption{Spinning signature in the TTAF for a single filament ($\Nf{=}1$, $\Mh{=}20$, $\Fact{=}20$).  (a)~$\kb{=}100$: oscillations decay by $\tau \approx 5000$.  (b)~$\kb{=}1000$: oscillations persist to $\tau \approx 15{,}000$.  Stiffer chains spin faster and more persistently.}
  \label{fig:ttaf_single}
\end{figure}

The TTAF independently characterizes the spinning state [Fig.~\ref{fig:ttaf_single}].  At $\kb = 100$ the TTAF oscillates with period $T_{\mathrm{spin}} \approx 80$--$90$, decaying by $\tau \approx 5000$; at $\kb = 1000$ the period shortens to ${\approx}\,60$ and coherence persists to $\tau \approx 15{,}000$.  Stiffer chains thus spin faster and more persistently.

\begin{figure*}[t]
  \centering
  \includegraphics[width=0.9\textwidth]{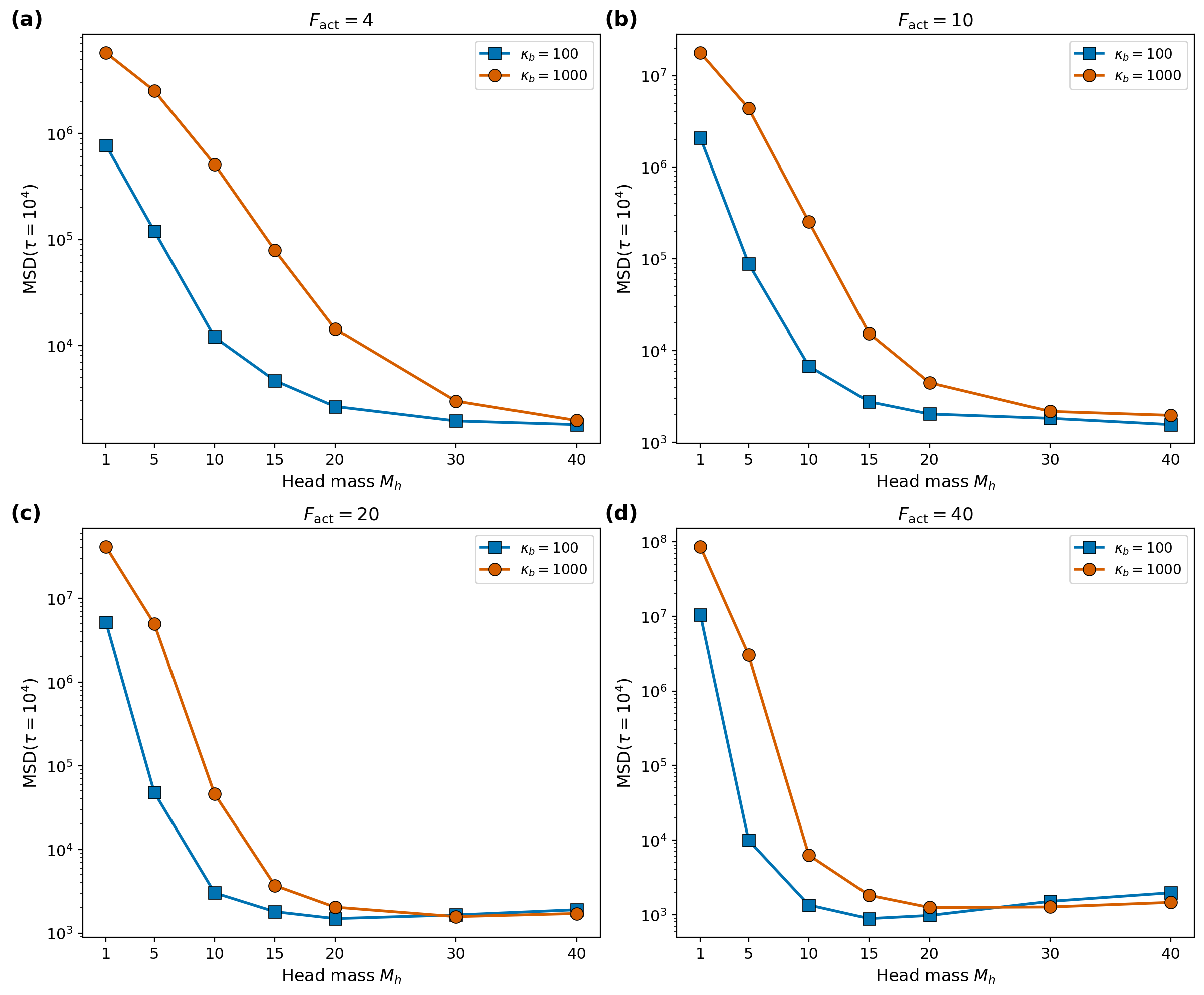} 
  \caption{Transport suppression in single filaments ($\Nf{=}1$): $\mathrm{MSD}_{\mathrm{ref}}$ vs.\ $\Mh$ at $\kb = 100$ and $1000$ for $\Fact = 4,\,10,\,20,\,40$.  Three to four orders of magnitude drop from light to heavy heads.}
  \label{fig:msd_ref}
\end{figure*}

Transport is severely suppressed: $\mathrm{MSD}_{\mathrm{ref}}$ drops by three to five orders of magnitude from $\Mh = 1$ to $\Mh = 20$ depending on $\kb$ and $\Fact$ (Fig.~\ref{fig:msd_ref}).  This motivates the central question: \emph{can multi-filament architecture overcome this arrest?}

\section{Multi-filament rescue}
\label{sec:rescue}

\begin{figure*}[t]
  \centering
  \includegraphics[width=\textwidth]{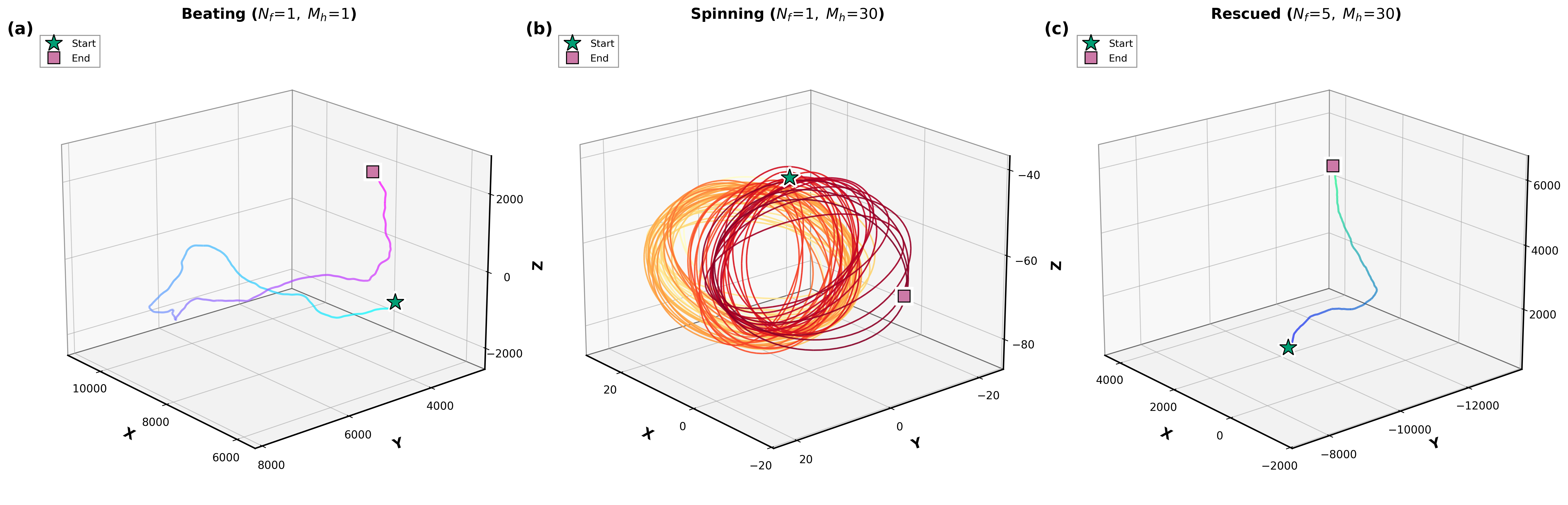}
  \caption{Three-dimensional head-bead trajectories illustrating the progression from beating to spinning arrest to multi-filament rescue (all at $\kb{=}1000$, $\Fact{=}40$).  (a)~Beating regime ($\Nf{=}1$, $\Mh{=}1$): the light head enables directed transport spanning thousands of $\sigma$.  (b)~Spinning arrest ($\Nf{=}1$, $\Mh{=}30$): confinement to ${\sim}30\sigma$.  (c)~Multi-filament rescue ($\Nf{=}5$, $\Mh{=}30$): directed transport comparable to the beating state is restored despite the heavy head.  Color encodes time progression; green star: start, pink square: end.}
  \label{fig:trajectories}
\end{figure*}

\subsection{Transport rescue: MSD recovery}
\label{sec:msd_rescue}

\begin{figure}[t]
  \centering
  \includegraphics[width=0.9\columnwidth]{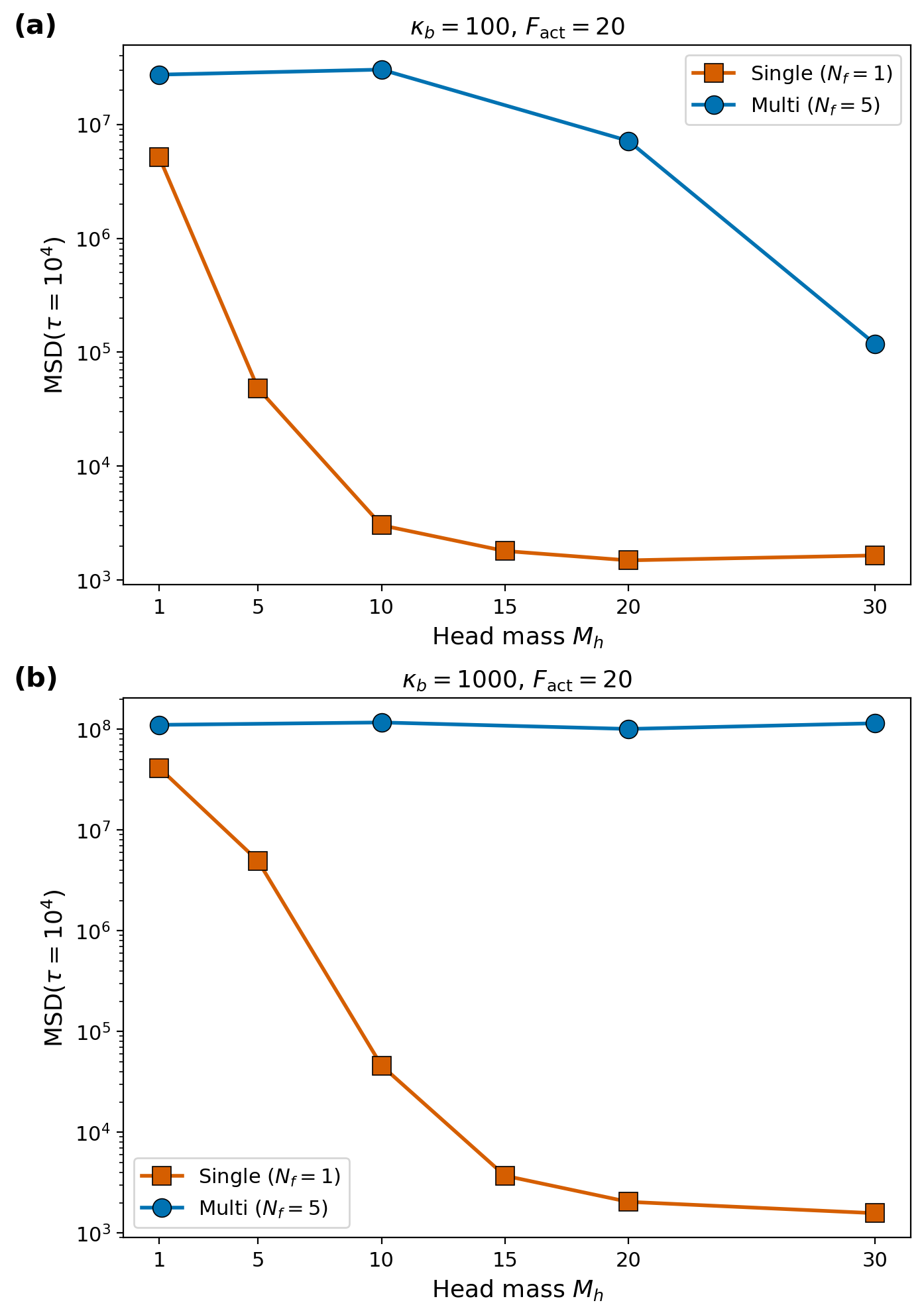}
  \caption{Transport rescue: $\mathrm{MSD}_{\mathrm{ref}}$ vs.\ $\Mh$ for single-filament ($\Nf{=}1$, orange) and five-filament ($\Nf{=}5$, blue) systems at $\Fact{=}20$.  (a)~$\kb{=}100$: two to three orders of magnitude rescue.  (b)~$\kb{=}1000$: five orders of magnitude rescue at $\Mh{=}30$.}
  \label{fig:rescue_msd}
\end{figure}

The visual contrast is immediate: three-dimensional trajectories (Fig.~\ref{fig:trajectories}) show that the spinning-arrested single filament at $\Mh{=}30$ is confined to ${\sim}30\sigma$, while the five-filament system recovers directed transport comparable to the unimpeded beating state.

Figure~\ref{fig:rescue_msd} quantifies the rescue at $\Fact = 20$.  At $\kb = 100$ [Fig.~\ref{fig:rescue_msd}(a)], the single-filament $\mathrm{MSD}_{\mathrm{ref}}$ drops from ${\sim}\,5 \times 10^4$ to ${\sim}\,10^3$ between $\Mh{=}1$ and $\Mh{=}20$, while the multi-filament system maintains $\mathrm{MSD}_{\mathrm{ref}} \approx 2 \times 10^7$ across this range---two to three orders of magnitude enhancement.  At $\Mh{=}30$ the rescue weakens, indicating limits at low stiffness.  At $\kb = 1000$ [Fig.~\ref{fig:rescue_msd}(b)], the rescue is dramatically stronger: the single filament collapses by a factor of $50{,}000$ while the multi-filament system holds at ${\sim}\,10^8$ even at $\Mh{=}40$, yielding approximately \emph{five orders of magnitude} enhancement at $\Mh{=}30$.

A striking result is the \emph{qualitative inversion of the role of stiffness}: $\kb = 1000$ produces worse single-filament transport but better multi-filament transport than $\kb = 100$, because the same rigidity that stabilizes the coiled spinning state in isolation promotes aligned bundled propulsion in the multi-filament system.

\subsection{Conformational rescue: SACF}
\label{sec:sacf_results}

\begin{figure}[t]
  \centering
  \includegraphics[width=0.9\columnwidth]{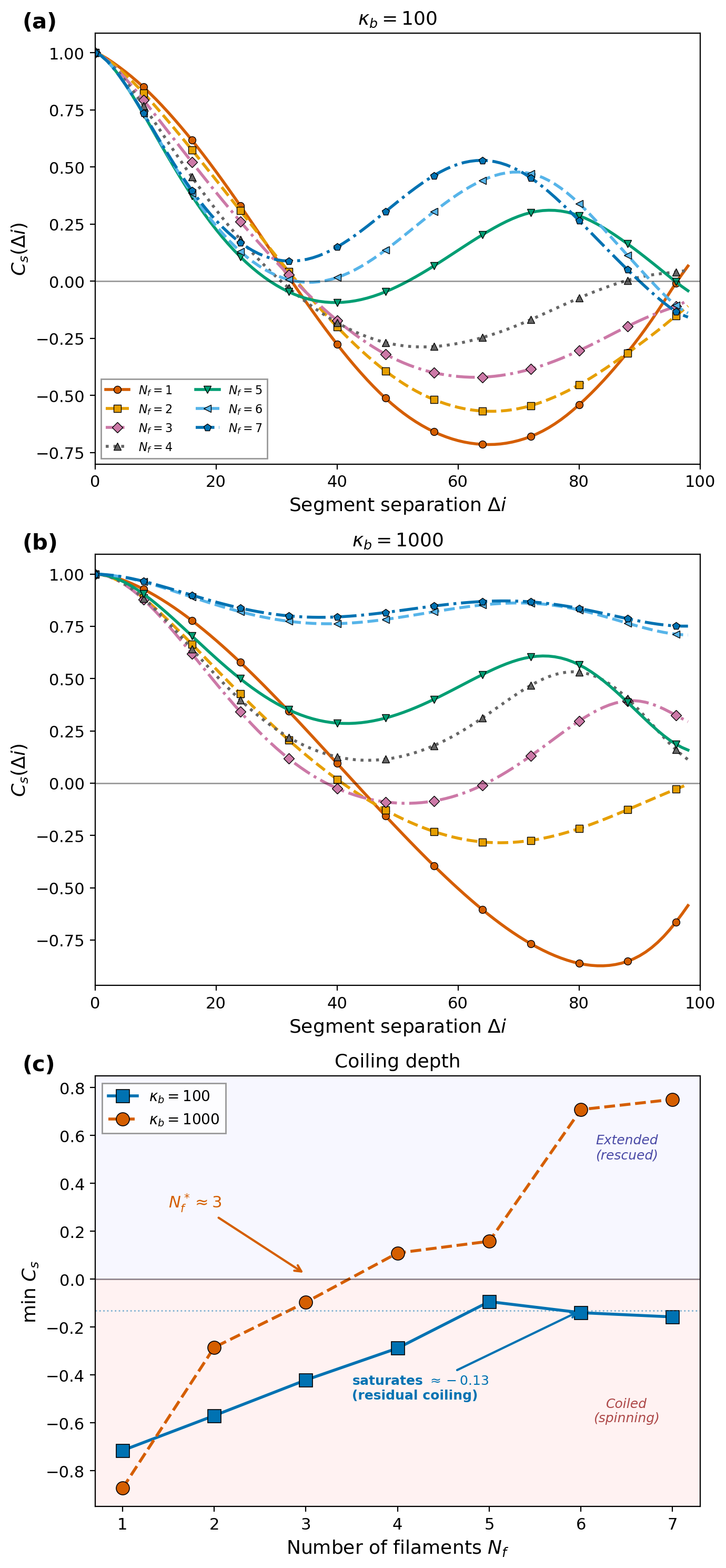}
  \caption{Conformational rescue.  SACF for selected $\Nf$ at $\Mh{=}20$, $\Fact{=}20$.  (a)~$\kb{=}100$: coiling depth progressively reduced but saturates near $\min\Cs \approx -0.13$.  (b)~$\kb{=}1000$: negative dip eliminated by $\Nf{=}3$; at $\Nf{\geq}6$ the SACF stays above $0.7$.  (c)~Minimum SACF value vs.\ $\Nf$: complete rescue (zero crossing) at $\kb{=}1000$ but saturation with residual coiling at $\kb{=}100$.}
  \label{fig:sacf}
\end{figure}

The SACF reveals the structural basis of the rescue.  For a single filament at $\Mh{=}20$, the SACF dips to $\min\Cs \approx -0.75$ at both stiffnesses [Fig.~\ref{fig:sacf}(a,b)], confirming strong helical coiling.  As $\Nf$ increases, this negative dip is progressively suppressed, but the two stiffnesses behave qualitatively differently [Fig.~\ref{fig:sacf}(c)].

At $\kb{=}1000$, the minimum SACF crosses zero near $\Nf^* \approx 3$ and rises to $+0.75$ by $\Nf{=}7$---the chains form a rigid aligned bundle with coiling fully eliminated.  At $\kb{=}100$, the coiling depth is reduced by more than $80\%$ but \emph{saturates} near $\min\Cs \approx -0.13$ for $\Nf \geq 5$, never crossing zero.  Flexible filaments retain sufficient conformational freedom to partially coil even when sterically constrained, whereas stiff filaments are geometrically incompatible with simultaneous coiling.

This asymmetry has an important consequence: at $\kb{=}100$, the transport rescue is already complete by $\Nf \approx 5$ (Sec.~\ref{sec:msd_rescue}) \emph{despite} the persistence of residual coiling.  The conformational and transport rescue thus decouple at low stiffness but coincide at high stiffness---a distinction whose physical origin is revealed by the TTAF analysis below.  The underlying mechanism is steric frustration: multiple filaments anchored to the same head cannot all simultaneously adopt the coiled conformation required for spinning, and the excluded-volume interactions force asymmetric configurations that generate net translational thrust.

\subsection{Orientational rescue: TTAF}
\label{sec:ttaf_results}

\begin{figure*}[t]
  \centering
  \includegraphics[width=0.9\textwidth]{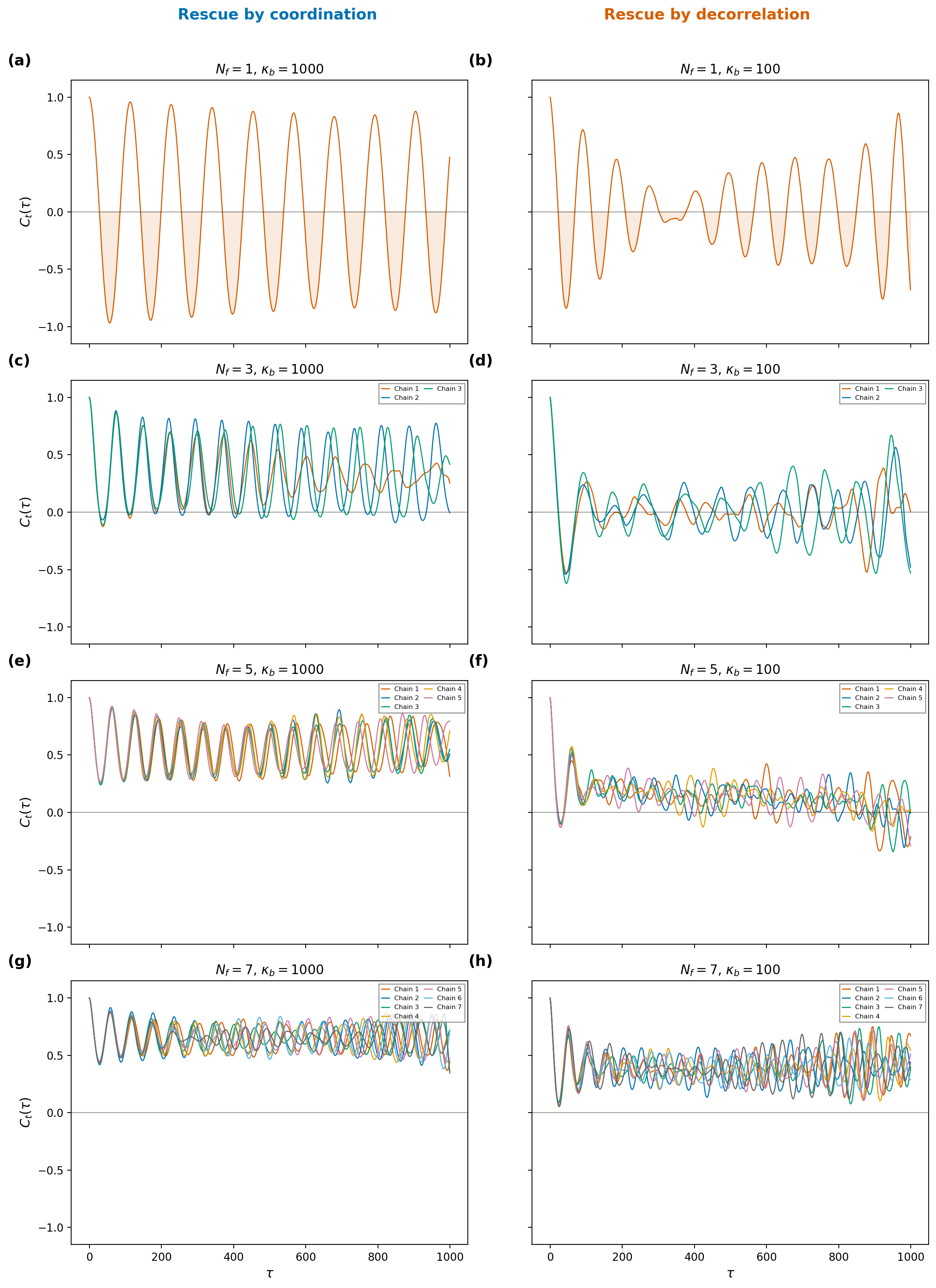}
  \caption{Orientational rescue: short-time TTAF ($\Mh{=}20$, $\Fact{=}20$).  Top row ($\kb{=}1000$, coordination): (a)~$\Nf{=}1$, full spinning; (b)~$\Nf{=}3$, entirely positive oscillations; (c)~$\Nf{=}5$, tight synchronization; (d)~$\Nf{=}7$, rigid bundle.  Bottom row ($\kb{=}100$, decorrelation): (e)~$\Nf{=}1$, decaying spinning; (f)~$\Nf{=}3$, rapid decoherence; (g)~$\Nf{=}5$, uncorrelated fluctuations; (h)~$\Nf{=}7$, disordered diffusion.  Individual chains in distinct colors.}
  \label{fig:ttaf_rescue}
\end{figure*}

The TTAF reveals two qualitatively distinct rescue mechanisms (Fig.~\ref{fig:ttaf_rescue}).

\paragraph{Rescue by coordination ($\kb{=}1000$, top row).}
For $\Nf{=}1$ the TTAF oscillates between $-1$ and $+1$ with period $T_{\mathrm{spin}} \approx 60$.  At $\Nf{=}3$, oscillations remain but stay entirely positive---the orientation never reverses, consistent with the SACF critical number $\Nf^* \approx 3$.  At $\Nf{=}5$ and $7$, all chains synchronize at high positive values, forming a coordinated bundle that sustains directed propulsion over timescales far exceeding $\tau_\mathrm{spin}$.

\paragraph{Rescue by decorrelation ($\kb{=}100$, bottom row).}
For $\Nf{=}1$, spinning oscillations decay by $\tau \approx 300$.  At $\Nf{=}3$, chains rapidly lose coherence, and for $\Nf \geq 5$ uncorrelated low-amplitude fluctuations persist throughout.  Spinning is eliminated but replaced by uncorrelated active diffusion rather than coordinated propulsion.

\paragraph{Two mechanisms, two outcomes.}
At $\kb{=}1000$, coordinated bundled propulsion achieves five orders of magnitude rescue; at $\kb{=}100$, decorrelated diffusion recovers only two to three orders.  Comparing panels~(c) and~(g)---both at $\Nf{=}5$---illustrates this directly: the same architectural fix produces tightly correlated chains ($C_t \approx 0.5$--$0.9$) in one case and disordered fluctuations near zero in the other.

This contrast also resolves the puzzle from Sec.~\ref{sec:sacf_results}: transport is rescued at $\kb{=}100$ despite residual coiling ($\min\Cs \approx -0.13$) because the steric interactions destroy the \emph{temporal coherence} of spinning without eliminating coiling per se.  Without coherent periodic reversal, the residual coiling produces only incoherent fluctuations, and the net thrust from $\Nf$ uncorrelated filaments drives enhanced diffusion.  It is the destruction of \emph{spinning coherence} that rescues transport in the flexible regime.

\subsection{Multi-filament conformational regimes}
\label{sec:conformations_multi}

\begin{figure}[t]
    \centering
    \begin{subfigure}[b]{0.45\columnwidth}
        \centering
        \includegraphics[height=1.5cm]{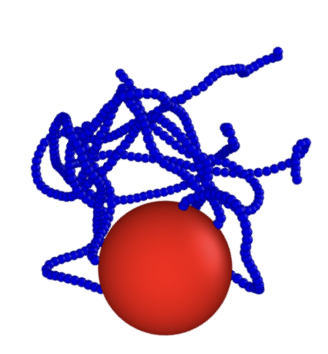}
        \caption{Disordered ($\kb{=}10$, $\Fact{=}20$)}
        \label{fig:conf_disordered}
    \end{subfigure}
    \hfill
    \begin{subfigure}[b]{0.45\columnwidth}
        \centering
        \includegraphics[height=3.5cm]{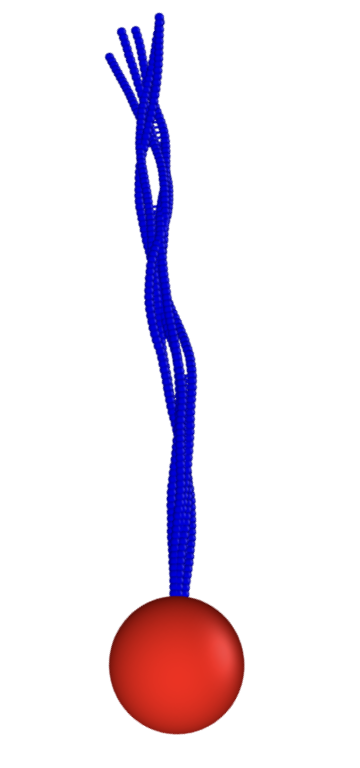}
        \caption{Bundled ($\kb{=}1000$, $\Fact{=}4$)}
        \label{fig:conf_bundled}
    \end{subfigure}
    
    \vspace{4pt}
    
    \begin{subfigure}[b]{0.45\columnwidth}
        \centering
        \includegraphics[height=3.5cm]{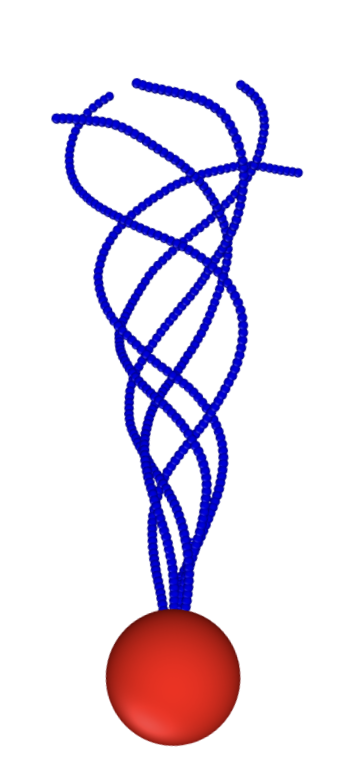}
        \caption{Helical ($\kb{=}1000$, $\Fact{=}20$)}
        \label{fig:conf_helical}
    \end{subfigure}
    
    \caption{Multi-filament conformational regimes (all at $\Nf{=}5$, $\Mh{=}20$).  (a)~Disordered: flexible filaments splay in random directions with residual local coiling (decorrelation pathway).  (b)~Bundled: stiff filaments at low activity form a tight, nearly parallel bundle (coordination pathway).  (c)~Helical: moderate activity at high stiffness introduces helical winding around the bundle axis.  At still higher activity ($\Fact{=}40$, not shown), the braiding intensifies further.  Red sphere: head bead; blue chains: active filaments.}
    \label{fig:conformations}
\end{figure}

The multi-filament assembly exhibits distinct conformational regimes as a function of $\kb$ and $\Fact$ (Fig.~\ref{fig:conformations}).  These arise from the competition between bending rigidity, active forcing, and steric interactions, and connect directly to the rescue mechanisms identified above.

In the \emph{disordered} regime [Fig.~\ref{fig:conformations}(a)], the flexible filaments splay out from the head in random directions with significant local curvature and entanglement.  Individual chains retain residual local coiling ($\min\Cs \approx -0.13$), visible as segments that double back on themselves, but the filaments are not organized into any coherent structure.  This regime corresponds to the decorrelation rescue pathway: steric interactions destroy spinning coherence without imposing inter-filament alignment.

In the \emph{bundled} regime [Fig.~\ref{fig:conformations}(b)], the stiff filaments at low activity form a tight, nearly parallel bundle reminiscent of a bacterial flagellar bundle.  The rigid backbones cannot coil around the head because neighboring filaments sterically prevent the required curvature.  This regime produces the strongest rescue and corresponds to the coordination pathway, where the SACF remains uniformly high ($> 0.7$) and the TTAF maintains long-lived positive correlations.  All filaments contribute thrust along a common direction, sustaining directed propulsion along the bundle axis.

In the \emph{helical} regime [Fig.~\ref{fig:conformations}(c)], moderate activity at high stiffness drives helical winding around the bundle axis while maintaining alignment near the head.  This state still produces effective transport recovery.  At still higher activity ($\Fact{=}40$, not shown), the braiding intensifies and the distal ends splay apart, reducing propulsion efficiency.  The progression from bundle to helix to braid suggests an optimal activity range for multi-filament propulsion: strong enough to generate thrust but not so strong as to destabilize the aligned bundle.

\section{Discussion and conclusions}
\label{sec:discussion}

We have demonstrated that multi-filament architectures rescue active transport from inertia-induced spinning arrest through steric frustration of the coiled conformation.  Three independent diagnostics---MSD, SACF, and TTAF---reveal two distinct rescue mechanisms: rescue by coordination (high stiffness, persistent bundled propulsion) and rescue by decorrelation (low stiffness, elimination of orientational coherence).  At $\kb{=}1000$, conformational and transport rescue coincide at $\Nf^* \approx 3$: coiling is fully eliminated and replaced by a rigid bundle.  At $\kb{=}100$, the conformational rescue is \emph{incomplete}---residual coiling persists ($\min\Cs \approx -0.13$) even at $\Nf = 7$---yet transport is still rescued by two to three orders of magnitude.  This decoupling demonstrates that disrupting the \emph{coherence} of spinning, not eliminating coiling per se, is sufficient for motility rescue.

\emph{Relation to prior work.}---The spinning arrest is consistent with spiral states reported by Karan \textit{et al.}~\cite{Karan2024} and Isele-Holder \textit{et al.}~\cite{IseleHolder2015}, extended here to three dimensions with an explicit heavy head.  The closest precedent for the rescue is the density-dependent spiral destabilization by Janzen and Matoz-Fernandez~\cite{Janzen2024}; our mechanism is qualitatively different, operating through the architectural constraint of the star topology for a single isolated molecule.  The active star polymer work of Buglakov \textit{et al.}~\cite{Buglakov2025} shares the multi-arm topology but did not address spinning suppression or inertia.

\emph{Steric frustration, not force accumulation.}---The rescue cannot be attributed to trivially having $\Nf$ times more active force.  Force scaling ($\mathrm{MSD}_\mathrm{ref} \propto \Nf^2$) accounts for at most $1.5$ orders of magnitude ($\Nf^2 = 25$ for $\Nf{=}5$), far short of the observed five orders.  Moreover, the phase diagram (Fig.~\ref{fig:phase_diagram}) shows that increasing activity on a \emph{single} filament \emph{deepens} the spinning trap---the multi-filament rescue operates against the force trend.  The conformational evidence (elimination of the SACF negative dip and cessation of TTAF zero crossings) confirms that the motility mode itself has changed, not merely the thrust magnitude.

\emph{Applicability and limitations.}---Our framework, excluding hydrodynamic interactions, is directly applicable to dry and quasi-dry settings: \emph{in vitro} motility assays~\cite{Schaller2010,Sumino2012,Jiang2014}, macroscopic active chains~\cite{Fazelzadeh2023,Deblais2020}, gliding bacteria~\cite{Nan2014}, and the hydrodynamically screened intracellular environment~\cite{Brangwynne2008}.  For multi-filament systems in fluid, HI would add coupling that mediates flagellar synchronization~\cite{Reigh2012,Brumley2012} and bundling~\cite{Kim2003,Flores2005}; our results establish the baseline that excluded-volume interactions alone achieve.  Additional limitations include equal filament lengths and fixed anchor geometry.

\emph{Outlook.}---The two rescue mechanisms suggest design principles for synthetic microswimmers~\cite{Palagi2018,Tottori2012,Li2017,Ceylan2019}: stiff filaments for directed transport, flexible filaments for enhanced diffusion.  Future work should incorporate HI, investigate the universality of the residual coiling saturation ($\min\Cs \approx -0.13$), and pursue experimental realization using motor-driven multi-filament assemblies or synthetic multi-flagellated swimmers.

\clearpage
\begin{acknowledgments}
A.B.\ acknowledges support from DST India (CRG/2023/000636), DBT India (BT/PR46247/BID/7/1015/2023), and the DBT Centre of Excellence grant.  A.D.\ acknowledges support from DST INSPIRE (DST/INSPIRE/04/2020/000928).
\end{acknowledgments}

\appendix
\section{Non-dimensionalization details}
\label{app:nondim}

We provide the complete non-dimensionalization of Eq.~\eqref{eq:langevin}.  The characteristic scales are: length $\sigma$, time $\td = \gamma\sigma^2/\kB T$, energy $\kB T$, and force $\kB T/\sigma$.

Substituting $\tilde{\vr}_i = \vr_i/\sigma$ and $\tilde{t} = t/\td$ into Eq.~\eqref{eq:langevin} gives
\begin{equation}
m_i\frac{\sigma}{\td^2}\,\ddot{\tilde{\vr}}_i = -\gamma_i\frac{\sigma}{\td}\,\dot{\tilde{\vr}}_i - \nabla_i U + f_a\,\vt_i + \sqrt{2\gamma_i\kB T}\;\bm{\eta}_i(t).
\end{equation}
Multiplying through by $\sigma/\kB T$ and using $\td = \gamma\sigma^2/\kB T$, each prefactor simplifies:

\emph{Inertial:}
\begin{equation}
\frac{m_i\sigma^2}{\kB T\td^2} = \frac{m_i\sigma^2}{\kB T} \cdot \frac{(\kB T)^2}{\gamma^2\sigma^4} = \frac{m_i\kB T}{\gamma^2\sigma^2} \equiv \mathcal{M}_i.
\end{equation}

\emph{Friction:}
\begin{equation}
\frac{\gamma_i\sigma^2}{\kB T\td} = \frac{\gamma_i\sigma^2}{\kB T} \cdot \frac{\kB T}{\gamma\sigma^2} = \frac{\gamma_i}{\gamma}.
\end{equation}

\emph{Active:} $\sigma f_a/\kB T = \Pe$.

\emph{Noise:} Defining the dimensionless noise $\tilde{\bm{\eta}}_i(\tilde{t}) = \sqrt{\td}\;\bm{\eta}_i(t)$, which satisfies $\langle\tilde{\eta}_i^\alpha(\tilde{t})\,\tilde{\eta}_j^\beta(\tilde{t}')\rangle = \delta_{ij}\delta_{\alpha\beta}\delta(\tilde{t} - \tilde{t}')$, the noise prefactor becomes $\sqrt{2\gamma_i/\gamma}$.

For body beads ($m_i = m$, $\gamma_i = \gamma$): $\mathcal{M}_i = \mathcal{M}$, $\gamma_i/\gamma = 1$, and the noise prefactor is $\sqrt{2}$, yielding Eq.~\eqref{eq:nondim}.

For the head bead ($m_\mathrm{h} = \alpha^3 m$, $\gamma_\mathrm{h} = \alpha\gamma$):
\begin{equation}
\Mh \equiv \frac{m_\mathrm{h}\kB T}{\gamma_\mathrm{h}^2\sigma^2} = \frac{\alpha^3 m\kB T}{(\alpha\gamma)^2\sigma^2} = \frac{\alpha^3}{\alpha^2}\cdot\frac{m\kB T}{\gamma^2\sigma^2} = \alpha\mathcal{M},
\end{equation}
the friction prefactor is $\gamma_\mathrm{h}/\gamma = \alpha$, and the noise prefactor is $\sqrt{2\alpha}$.  The dimensionless head equation is therefore
\begin{equation}
\alpha\mathcal{M}\,\ddot{\tilde{\vr}}_\mathrm{h} = -\alpha\,\dot{\tilde{\vr}}_\mathrm{h} - \tilde{\nabla}_\mathrm{h}\tilde{U} + \Pe\,\vt_\mathrm{h} + \sqrt{2\alpha}\;\tilde{\bm{\eta}}_\mathrm{h},
\end{equation}
which is Eq.~\eqref{eq:head} of the main text (with forces summed over all filaments for the multi-filament case).  The linear scaling $\Mh = \alpha\mathcal{M}$ arises because the head mass scales as $\alpha^3$ while $\gamma_\mathrm{h}^2$ scales as $\alpha^2$, leaving a single power of $\alpha$.

\emph{Physical interpretation.}---The three dimensionless parameters separate cleanly: $\mathcal{M} = \ti/\td$ controls inertia (independently of activity), $\Pe = \sigma f_a/\kB T$ controls activity (independently of mass), and $\alpha$ controls geometry.  This allows systematic studies where, e.g., $\Pe$ is fixed while $\mathcal{M}$ is varied to isolate inertial effects, or vice versa.

\emph{Comparison with the active-time scheme.}---An alternative non-dimensionalization uses $\ta = \sigma/v_0$ as the time reference, yielding the dimensionless mass $\mathcal{M}' = \ti/\ta = m v_0/(\gamma\sigma) = mf_a/(\gamma^2\sigma)$.  This definition couples inertia and activity in a single parameter: changing $f_a$ changes $\mathcal{M}'$ even at fixed mass.  The head mass becomes $\Mh' = m_\mathrm{h}v_0/(\gamma_\mathrm{h}\sigma) = \alpha^2 m v_0/(\gamma\sigma) = \alpha^2\mathcal{M}'$, with a stronger geometric coupling.  Table~\ref{tab:nondim} summarizes the comparison.

\begin{table}[h]
\caption{Comparison of non-dimensionalization schemes.}
\label{tab:nondim}
\begin{ruledtabular}
\begin{tabular}{lcc}
Aspect & $\mathcal{M}' = \ti/\ta$ & $\mathcal{M} = \ti/\td$ (this work) \\
\hline
Reference time & $\sigma/v_0$ & $\gamma\sigma^2/\kB T$ \\
$\mathcal{M}$ definition & $m f_a/\gamma^2\sigma$ & $m\kB T/\gamma^2\sigma^2$ \\
Activity dependence & Yes & No \\
$\Mh$ & $\alpha^2\mathcal{M}'$ & $\alpha\mathcal{M}$ \\
Parameter separation & Mixed & Clean \\
\end{tabular}
\end{ruledtabular}
\end{table}


\end{document}